# Localized heat diffusion in topological thermal materials

Minghong Qi[1,2,†], Dong Wang[1,2,†], Pei-Chao Cao[3], Xue-Feng Zhu[3], Cheng-Wei Qiu[4]*, Hongsheng Chen[1,2]*, & Ying Li[1,2]*

[1] *Interdisciplinary Center for Quantum Information, State Key Laboratory of Modern Optical Instrumentation, ZJU-Hangzhou Global Scientific and Technological Innovation Center, Zhejiang University, Hangzhou 310027, China*

[2] *International Joint Innovation Center, Key Lab. of Advanced Micro/Nano Electronic Devices & Smart Systems of Zhejiang, The Electromagnetics Academy of Zhejiang University, Zhejiang University, Haining 314400, China*

[3] *School of Physics and Innovation Institute, Huazhong University of Science and Technology, Wuhan, 430074, China*

[4] *Department of Electrical and Computer Engineering, National University of Singapore, Singapore 117583, Singapore*

† These authors contributed equally to this work.

*e-mail: chengwei.qiu@nus.edu.sg; *e-mail: hansomchen@zju.edu.cn; * e-mail: eleying@zju.edu.cn

**Various unusual behaviors of artificial materials are governed by their topological properties, among which the edge state at the boundary of a photonic or phononic lattice has been captivated as a popular notion. However, this remarkable bulk-boundary correspondence and the related phenomena are missing in thermal materials. One reason is that heat diffusion is described in a non-Hermitian framework because of its dissipative nature. The other is that the relevant temperature field is mostly composed of modes that extend over wide ranges, making it difficult to be rendered within the tight-binding theory as commonly employed in wave physics. Here, we overcome the above challenges and perform systematic studies on heat diffusion in thermal lattices. Based on a continuum model, we introduce a state vector to link the Zak phase with the existence of the edge state, and thereby analytically prove the thermal bulk-boundary correspondence. We experimentally demonstrate the predicted edge states with a topologically protected and localized heat dissipation capacity. Our finding sets up a solid foundation to explore the topology in novel heat transfer manipulations.**

Heat transfer plays an important role in science and engineering, such as energy management[1,2], and thermal transport[3], radiative cooling[4], etc. The related parameters, such as thermal conductivity, density, heat capacity, and Seebeck coefficient, are natively constrained by the material in use. A lot of investigations have been thus dedicated to the advanced manipulation of heat transfer with artificial structures. For example, thermal metamaterials[5,6] can realize many unique and useful effects such as thermal cloaking[7,8], thermal rectification[9,10], infrared camouflage[11,12,13], and continuous tunablility[14,15], primarily at steady state. On the other hand, the transient heat transfer has been largely neglected, because a robust and precise framework to theoretically describe the transient performance of thermal metamaterials is yet to be established. It prevents us, intuitively and methodologically, from accessing the unknown behaviors, novel physics, and far-reaching applications in heat transfer.

Recently, several works point out that an effective Hamiltonian can be employed to describe and predict the transient temperature field evolution[16,17,18]. Owing to the dissipative nature of heat transfer, its effective Hamiltonian turns out to be non-Hermitian. There have been plenty of studies on non-Hermitian wave systems, uncovering their unusual symmetric and topological properties[19-23]. For example, effects like unidirectional transparency[24], coherent perfect absorption[25], and single-mode lasing[26] can be realized by using parity-time (PT) symmetry. A related concept, anti-parity-time (APT) symmetry, has been demonstrated in convective heat transfer with a locking effect on the hot spot[16,17].

However, there are still intrinsic difficulties in exploring the transient behaviors in pure heat diffusion. As shown in this work, the effective Hamiltionian for heat diffusion is anti-Hermitian in a suitable Hilbert space. Its eigenvalues are pure imaginary numbers, representing the decay rate of each eigenmode. It follows that there is no resonance for a diffusive system, due to the lack of a driving source at imaginary frequency. It is almost impossible to selectively excite and study a diffusion mode, which is in big contrast to the wave cases (Fig. 1a). Generally, the transient heat diffusion is as shown in Fig. 1b, where a local hot spot gradually spreads out and decays. Such a field is composed of multiple modes, among which those with small decay rates that extends over wide ranges are eventually observed. Because the observable field is mostly composed of extended modes, it is very difficult to confine a diffusion mode in a thermal material as in a wave resonator. Therefore, field localization is rare in heat diffusion, except for the trivial case in a well insulated part. A localized heat diffusion as in Fig. 1c is helpful for the heat dissipation of a hot spot without influencing its neighborhoods, but its realization and physical mechanisms are still elusive.

Topological edge state is a kind of localized mode originated from the nontrivial band topology of a lattice structure. This bulk-boundary correspondence is a remarkable finding demonstrated in various wave systems[27], which might be the awaited mechanism to localize a diffusion mode. However, the bulk-boundary correspondence has never been experimentally demonstrated in thermal materials yet, due to the sharp contrast between diffusion and wave dynamics.

Here, for the first time, we rigorously prove the bulk-boundary correspondence in heat diffusion with a continuum model, and experimentally observe the localized heat diffusion protected by the edge state. The band structure, geometric phase, and edge states of the one-dimensional heat transfer lattice are analytically solved to reveal their close relationships. We also predict and demonstrate different topological phases controlled by the boundary condition. Our results show that the edge states have a topologically protected heat dissipation capability, which may help in the cooling of local hot spots.

**One-dimensional thermal lattice**

We present a 3D model composed of spheres and rods with the same density $\rho$, heat capacity $c$, and thermal conductivity $\kappa$ as shown in Fig. 1d. The lengths occupied by the two spheres of radius $R$ in one unit are $b_1$ and $b_3$ (= $b_1$). The lengths of the intra-cell and inter-cell rods with radius $R_0$ are $b_2$ and $b_4$. We define a parameter $c$ to represent the length difference between the two rods, i.e., $b_4 - b_2 = 2c$. Rod lengths can be tailored to control heat exchange efficiencies within and between the units. The system intuitively seems to follow the Su-Schrieffer-Heeger (SSH) model[28].

However, we shall show that the system cannot be described rigorously by using the discrete Hamiltonian of the SSH model.

To rigorously analyze the model, we consider a one-dimensional equivalent model. The region of the sphere is modeled as a rod with equivalent heat capacity $c_{p1}$, density $\rho_1$, and thermal conductivity $\kappa_1$. The equivalent heat capacity multiplied by the density is estimated by the average volume ratio, $\rho_1 c_{p1} = \rho c_p V_{\text{sphere}}/(\pi R_0^2 b_1)$, where $V_{\text{sphere}}$ is the volume of the sphere part (the grey region in Fig. 1d). The equivalent thermal conductivity is estimated by the average cross-sectional area ratio, $\kappa_1 = \kappa S_{\text{sphere}}/(\pi R_0^2)$, where $S_{\text{sphere}}$ is the average cross-sectional area of the region occupied by a sphere. The region of the rod keeps the same as the 3D model with equivalent heat capacity $c_{p2}$, density $\rho_2$, and thermal conductivity $\kappa_2$.

We define the middle of the intra-cell rod ($x = 0$) as the center of a unit. The 1D governing equation of the temperature field $T(x,t)$ is

$$\rho c_p(x) \frac{\partial T}{\partial t} = \frac{\partial}{\partial x}\left( \kappa(x) \frac{\partial T}{\partial x} \right) \tag{1}$$

where $T$ is the temperature, $x$ is the direction of the chain model, $t$ is the time. $\rho c_p(x) = \rho c_{pm}$ and $\kappa(x) = \kappa_m$ are piecewise functions, where $m = 1$ for the sphere-parts and $m = 2$ for the rod-parts.

Assuming that the solution to Eq. (1) has the form $T(x,t) = u(x)e^{-i\omega t}$, Eq. (1) becomes an eigenvalue problem

$$\omega u(x) = \frac{i}{\rho c_p(x)} \frac{\partial}{\partial x}\left( \kappa(x) \frac{\partial u}{\partial x} \right) = \mathbf{H} u(x) \tag{2}$$

where $\mathbf{H}$ is the effective Hamiltonian. If we define the inner product between two states $u$ and $v$ as

$$\langle v | u \rangle = \int \rho c_p(x) v^* u \, dx \tag{3}$$

where the integral is performed in a unit, the effective Hamiltonian is anti-Hermitian. Therefore, $\omega$ is a pure imaginary number $\omega = -i\lambda$, with $\lambda$ representing the decay rate.

Considering the periodicity of the system, the Bloch theorem gives $u(x) = \psi(x)e^{ikx}$, where $\psi(x)$ is periodic in a unit cell, and $k$ is the wavenumber. Using plane-wave expansion, $u(x)$ can be written as a piecewise function in the $n$-th unit ($-a/2 \leq x - na \leq a/2$)

$$u(x) = X^{(j)}_n e^{ik_m(x-na)} + Y^{(j)}_n e^{-ik_m(x-na)}, \zeta_j < x - na < \zeta_{j+1} \qquad (4)$$

where $k_m = (\lambda \rho_m c_{pm}/\kappa_m)^{1/2}$ are obtained from the dispersion relations. For $j = \pm 1$, $m = 1$. For $j = 0$ and $\pm 2$, $m = 2$. $\zeta_j$ represents the coordinate at the interface: $\zeta_{-2} = -a/2$, $\zeta_{-1} = -(a - b_4)/2$, $\zeta_0 = -b_2/2$, $\zeta_1 = b_2/2$, $\zeta_2 = (a - b_4)/2$, $\zeta_3 = a/2$. According to the matching conditions between regions in above equations, the transfer matrix[29] $M$ from unit $n - 1$ to unit $n$ is:

$$\begin{pmatrix} X^{(-2)}_n \\ Y^{(-2)}_n \end{pmatrix} = \begin{pmatrix} M_{11} & M_{12} \\ M_{21} & M_{22} \end{pmatrix} \begin{pmatrix} X^{(-2)}_{n-1} \\ Y^{(-2)}_{n-1} \end{pmatrix} = e^{ika} \begin{pmatrix} X^{(-2)}_{n-1} \\ Y^{(-2)}_{n-1} \end{pmatrix} \qquad (5)$$

where the Bloch theorem has been used. Thanks to the anti-Hermiticity of $H$, $k_{1,2}$ is real. It follows that $M_{11} = M_{22}^*$ and $M_{12} = M_{21}^*$. We also have $\det(M) = 1$.

Eq. (5) is an eigenvalue problem for $M$, which gives (see Supplementary Note 1 for detailed derivations)

$$\cos ka = \text{Re}(M_{11}) = 2\left(\cos\alpha \cos\beta - \cosh\gamma \sin\alpha \sin\beta\right)^2 - 2\sinh^2\gamma \sin^2\alpha \sin^2\delta - 1 \qquad (6)$$

where $\alpha = k_1 b_1$, $\beta = k_2(b_2 + b_4)/2$, $\delta = k_2(b_4 - b_2)/2 = k_2 c$, $\gamma = \ln(k_1\kappa_1/k_2\kappa_2) = [\ln(\kappa_1 \rho_1 c_{p1}/\kappa_2 \rho_2 c_{p2})]/2$. The band structure can be obtained by solving $\lambda$ from Eq. (6). The results are shown as red dots in Fig. 1f. When $c = 0$, the two bands are degenerate at $k = \pm\pi/a$. When $|c| > 0$, the right-hand-side of Eq. (6) satisfies $\text{Re}(M_{11}) < -1$, resulting in a bandgap opened at $k = \pm\pi/a$.

We perform finite element simulations using COMSOL Multiphysics, and the results are shown in Fig. 1f. The 1D band distribution solved from Eq. (6) is consistent with the 3D model numerical results. The slight differences are mainly due to the small errors introduced in the estimation of the material parameters of the 1D system. It can also be found that the eigenvalues of the discrete Hamiltonian matrix from the SSH model are not in agreement with the simulation results (see Supplementary Note 2).

**The Zak phase**

Generally, the topological characteristics of a non-Hermitian system are very different from those of the Hermitian system[30,31]. The non-Hermitian skin effect where the eigenstates in the bulk bands become localized boundary modes (non-Bloch modes) can significantly modify the original bulk-boundary correspondence in the Hermitian systems[32-36]. Fortunately, the anti-Hermiticity of our Hamiltonian ensures an ordinary Hilbert space, which supports a well defined geometric phase in our system.

The geometric phase is at the kernel of many topological systems. For 1D systems, the Zak phase[37,38,39] is such a parameter defined as:

$$\mathcal{Z} = \int_{-\pi/a}^{\pi/a} \langle \psi(x,k) | i\partial_k | \psi(x,k) \rangle \, dk \tag{7}$$

where $\psi(x,k)$ is normalized using the inner product defined in Eq. (3). For inversion symmetric unit cell, a state and its spatial inversion are related by a large gauge transformation[40] $\psi(-x,-k) = e^{i\phi(k)}\psi(x,k)$. When there is no singularity at $k = 0$, one can show that $\mathcal{Z} = \phi(\pi/a) - \phi(0)$. Therefore, the Zak phase is 0 when $u(x, \pm\pi/a)$ and $u(x, 0)$ have the same symmetry. Otherwise it is $\pi$. We note that there is some recent debate on the quantization of Zak phase[41]. However, that work is based on the analysis of Schödinger equation, which is different from the governing equation Eq. (1) in this work. The properties of $\psi(x,k)$ are thus different. For example, they cannot be gauged to pure real functions even in the presence of symmetric $\kappa(x)$ and $\rho c_p(x)$.

In our case, the eigenstate is determined by the eigenvector of the transfer matrix:

$$\begin{pmatrix} X^{(-2)} \\ Y^{(-2)} \end{pmatrix} \propto \begin{pmatrix} M_{12} \\ e^{ika} - M_{11} \end{pmatrix} \propto \begin{pmatrix} e^{ik_2a/2} \operatorname{Im}(M_{12}') \\ e^{-ik_2a/2}[\sin ka - \operatorname{Im}(M_{11})] \end{pmatrix} = \begin{pmatrix} e^{ik_2a/2} & 0 \\ 0 & e^{-ik_2a/2} \end{pmatrix} U \qquad (8)$$

For the sake of convenience, we define $M_{12}' = \exp(-ik_2a)M_{12}$, which is a purely imaginary number, and a state vector $U = (U_1, U_2)^T$ with two real entries. Note that $U_1$ and $U_2$ are the values of the forward and backward parts in $u(-a/2)$. Based on the transfer matrix, we can show the following properties of the state vector: (1) For the symmetric case $U_1 = U_2$ with respect to inversion at $x = -a/2$, $u(x,0)$ is even and $u(x,\pm\pi/a)$ is odd; (2) For the anti-symmetric case $U_1 = -U_2$, $u(x,0)$ is odd and $u(x, \pm\pi/a)$ is even; (3) The sign of $U_2$ will not change in a band. According to properties (1) and (2), the Zak phase of a band is $\pi$ or $0$ when the state vector has the same or different symmetry at $k = 0$ and $\pm\pi/a$. Then according to property (3), if a sign change of $U_1$ happens between $k = 0$ and $\pm\pi/a$, the Zak phase is $0$. Otherwise, it is $\pi$.

One exception is at $k = 0$ ($\lambda = 0$) in the lower band, where $U$ is unsmooth due to the square-root singularity on $\lambda$. A phase of $\pi$ will be added to the Zak phase without a sign change of $U_1$. Therefore, for the lower band the above relationship should be reversed. When the length difference $c < 0$, a sign change of $U_1$ happens in the lower band, so $\mathcal{Z}_1 = \pi$. When $c > 0$, no sign change of $U_1$ happens, so $\mathcal{Z}_1 = 0$. For the Zak phase $\mathcal{Z}_2$ of the upper band, we similarly verify that $\mathcal{Z}_2 = \pi$ for $c < 0$, while $\mathcal{Z}_2 = 0$ for $c > 0$. Therefore, the Zak phase $\mathcal{Z} = \mathcal{Z}_1 = \mathcal{Z}_2$ can be used as a global topological invariant for the system. Two distinct topological regimes are identified according to the opposite signs of $c$. Figure 2 shows the trajectories of $U$ for the lower and upper bands at $c = -8$ mm and 8 mm, respectively. They confirm all our theoretical predictions, especially the close relationship between the sign change of $U_1$ and the Zak phase. More detailed observations of the topological transition have been demonstrated in Supplementary Fig. 2 for the trajectories at $c = -1$, 0, and 1 mm.

**Bulk-boundary correspondence**

In topological systems, when the topological invariant is nontrivial, there will be edge states. A recent theoretical effort[42] has been made to describe heat diffusion with the SSH model, but the rigor or appropriateness of using tight-binding approximation in thermal materials is in question (see Supplementary Note 2), largely due to the non-locality of most diffusion modes. Therefore, the thermal bulk-boundary correspondence has not been demonstrated in a rigid fashion. Here we provide a rigorous proof based on our continuum model. Under the parameters used in Fig. 1e, we build a chain with $N + 1$ units and apply constant boundary conditions at its two ends. It is worth noting that the selection of different boundary lengths will greatly impact the edge states, which we will show later.

To construct a finite length chain, we consider two cases of boundaries. For Case 1, we take the center of the inter-cell rod as the boundary of the chain as shown in Fig. 3a. For Case 2, the rods at both ends are extended, where the length of each extended part is $b_5/2$. The general method is to take the boundary of the unit as the boundary of the chain, so we study Case 1 first.

At the two ends of the chain, we have

$$\begin{cases} X^{(-2)}{}_0 e^{-ik_2 a/2} + Y^{(-2)}{}_0 e^{ik_2 a/2} = 0 \\ X^{(2)}{}_N e^{ik_2 a/2} + Y^{(2)}{}_N e^{-ik_2 a/2} = 0 \end{cases} \quad (9)$$

Eq. (9) can be written in matrix form using the transfer matrix $M$. To simplify the result, note that the eigenvalue of the transfer matrix is $\Lambda_\pm = \text{Re}(M_{11}) \pm [\text{Re}(M_{11})^2 - 1]^{1/2}$. The concerned edge states we are concerned about should appear in the bandgap. In order to meet the condition, Eq. (6) must have no real solution for $k$, leading to $\text{Re}(M_{11}) < -1$. Therefore, $\Lambda_+$ and $\Lambda_-$ are real numbers. Therefore, it is possible to diagonalize the transfer matrix $M$ despite its non-Hermiticity. Thus, we can rewrite Eq. (9) as

$$\begin{pmatrix} e^{-ik_2a/2} & e^{ik_2a/2} \end{pmatrix} \begin{pmatrix} X_+^R & X_-^R \\ Y_+^R & Y_-^R \end{pmatrix} \begin{pmatrix} \Lambda_+^N & 0 \\ 0 & \Lambda_-^N \end{pmatrix} \begin{pmatrix} X_+^L & Y_+^L \\ X_-^L & Y_-^L \end{pmatrix} \begin{pmatrix} e^{ik_2a/2} \\ -e^{-ik_2a/2} \end{pmatrix} = 0 \qquad (10)$$

where $(X_-^R, Y_-^R)^T$ and $(X_-^L, Y_-^L)$ are the normalized right and left eigenvectors of the transfer matrix. Since $\text{Re}(M_{11}) < -1$, $-1 < \Lambda_+ < 0$. When $N$ approaches infinity, $\Lambda_+^N$ approaches zero, so we can obtain the location of the edge states as $\text{Re}(M_{11})^2 = 1$ and $\text{Im}(M_{11}) = \text{Im}(M_{12}') = U_1$. It follows that the eigenvalue is located at the band edge. We thus show that no edge state can be found for Case 1.

For Case 2, we have $[\text{Re}(M_{11})^2 - 1]^{1/2} = -U_1\sin k_2 b_5$ and $\text{Im}(M_{11}) = U_1\cos k_2 b_5$. It is easy to see that Case 1 is a special case when $b_5 = 0$. We also note that the two equations are not independent, but related by $\det(M) = 1$. Figure 3e shows the solutions for $b_5 = b_1 + b_4$, which are in good agreement with the numerical results.

Based on the requirement that $[\text{Re}(M_{11})^2 - 1]^{1/2} > 0$, we define a parameter as the indicator of edge state $\eta = U_1\sin k_2 b_5$, it follows that Eq. (10) only has a solution when $\eta < 0$. Whether the system has an edge state is thus determined by the sign of $\eta$. We note that $U_1$ should be nonzero in the bandgap, since $|M_{12}|^2 = \text{Re}(M_{11})^2 + \text{Im}(M_{11})^2 - 1 > 0$. Therefore, the sign of $U_1$ does not change inside the bandgap. It is thus completely determined by its sign at the upper edge of the lower band. Considering that $U_1 > 0$ in the vicinity of $\lambda = 0$ (when $\sinh\gamma > 0$ in our case) and the previous discussion on the Zak phase $\mathcal{Z}_1$ of the lower band, the relation $\text{sgn}(U_1) = \exp(i\mathcal{Z})$ can be obtained. Our system is thus proved to possess bulk-boundary correspondence, which depends on the extension length $b_5$. The dependence of $\eta$ on $b_5$ and $c$ is plotted in Fig. 3b. Edge state exists in regimes below the zero plane (see Supplementary Fig. 3 for the positions of the edge states at different $b_5$).

Similarly, we can show that the system with thermally insulated boundaries exhibit edge states when $\eta > 0$, so one can observe them for $c > 0$ with $\mathcal{Z} = 0$ (see Supplementary Note 3). This result for open boundary conditions is contrary to the prediction based on the SSH model[38], where edge states should exist when the intra-cell coupling is smaller than the inter-cell coupling ($c < 0$). This is because the actual modes are not localized on the spheres, but on the rods (see Supplementary Fig. 4), making the tight-binding model invalid. It again confirms the necessity to analyze the system based on our continuum model and the Zak phase. For various other boundary conditions, we can generalize the indicator $\eta$ to $U_1 \sin(k_2 b_5 + 2\theta)$, where $\theta$ characterizes the hybridization between fixed and open boundary conditions (see Supplementary Note 3 and Supplementary Fig. 5). Edge states can also be found at the interface between the model and its mirroring one, with insulative boundary conditions (see Supplementary Fig. 6 for the band structure and edge state field of the model).

To further optimize the performance of the model in the edge state, the bandgap interval needs to be enlarged, which can be achieved by reducing the intra-cell coupling strength. It is accomplished by adding holes to the intra-cell rods of the model when $c = -8$ mm as shown in Fig. 3d. As demonstrated in the eigenvalue map (Fig. 3e), the bandgap gradually increases along with the number of holes. Figure 3f shows the temperature field of two cases with $c = -8$ mm in the edge state ($\lambda = 5.17$ mHz). For Case 1, there is no edge state, and the temperature field of the model is periodically distributed.

**Experimental observation of localized heat diffusion**

In order to experimentally observe the edge states, we can measure and analyse the evolution of the temperature field in the system by under appropriate initial conditions. For convenience, the chain models are bent into rectangles, with metal plates connected to them at the bottoms (Fig. 4a). The temperature at the boundaries of the model is controlled by making contact between the metal plates and the constant temperature heat sinks. We use a small-caliber hot air gun to heat the 1st (edge) and 20th (bulk) spheres respectively. Here we record the time when the maximum temperature of the heated sphere drops to 320 K, which is defined as $t = 0$.

We measured the decay rate curves of the maximum temperature for the heated spheres over time. As shown in Fig. 4b, model I, II, and III respectively stand for models with four holes when $c = -8$ mm, no holes when $c = 0$ mm, and four holes when $c = 8$ mm. According to theoretical predictions, the eigenstate of the system should decay exponentially with time, and its decay rate $\lambda$ corresponds to the imaginary part of the eigenvalue.

The decay rate rate of model I. edge (red line) meets with the speed predicted by the eigenvalue of the edge state in this case (black dotted line). It indicates that the initial temperature field of a hot spot localized on the edge sphere of model I should be close to the edge state distribution in Fig. 3f, as confirmed by the measured results. For model III, it is theoretically predicted that there is no edge state, so the initial temperature field of model III. edge does not meet any eigenmode. Its decay rate varies with time and is the same as that of model I. bulk and III. bulk. The situation of model II is at the critical point between the trivial and non-trivial topological phases. The decay rate of model II. edge is greater than that model II. bulk, but it does not follow an exponential time dependence.

The temperature distribution of the three models at 40 s is shown in Fig. 4c, where the maximum temperature of model I. edge is the lowest. The initial hot spot on the edge sphere of model I also introduces the smallest heating to the adjacent sphere, demonstrating a well localized heat diffusion. Figure 4d shows the experimentally measured thermal images of the three models, where the hot parts are enlarged as insets. It can be found that the experimental results are in line with theoretical expectations. When the same initial temperature is applied, at the same time, model I. edge has the lowest maximum temperature with the most confined temperature field.

The simulation results for the three models could be found in Supplementary Fig. 7, which are consistent with the experimental results. For the model of a single sphere, it can be seen that the final boundary state will not spread out, and its maximum temperature will be lower than that of the single sphere model (Supplementary Fig. 8).

The full spectrum of the system can also be probed through multiple sets of experimental measurements. We define a column vector $\hat{T}(t) = (T_1, T_2, …, T_{40})^T$ to represent the temperature field (the room temperature is subtracted), with $T_n(t)$ referring to the temperature of the $n$-th sphere at time $t$. We perform 40 series of experiments. For the $n$-th experiment, an initial temperature $T_{in}$ is applied to the $n$-th sphere, so we have $\hat{T}(t) = e^{-i\hat{H}t}\hat{T}_n(0)$, where $\hat{H}$ is a 40 × 40 matrix discretized from the Hamiltonian $H$ in Eq. (2). Therefore, the eigenvalues of $\exp(-i\hat{H}t)$ should approximate $\exp(-\lambda t)$. In Fig. 4e, the experimental eigenvalues are compared with the numerical results, which are basically consistent. It can be directly observed that model I has boundary states while model III does not, which reflects different topological phases.

**Conclusion and discussions**

We provide the first rigorous proof of bulk-boundary correspondence and the first experimental observation of topological edge state in heat diffusion. A continuum model is proposed to describe the heat conduction in a thermal lattice and give accurate analytical solutions for the band structure. By calculating the Zak phase, we reveal the topological origin of the edge states and the complex dependence of their positions upon boundary conditions in the heat diffusion. We numerically verify and optimize the band structure, and experimentally observe the edge state that dissipates locally and exponentially at the predicted rate. The localized heat diffusion may be further explored to facilitate the design of novel thermal materials and devices for efficient and robust heat manipulation.

**Methods**

In the theory and simulations, the parameters used in this model are: sphere radius $R = 4$ mm, rod radius $R_0 = 2.2$ mm, unit length $a = 32$ mm, material parameter density $\rho = 8000$ kg/m$^3$, heat capacity $c_p = 500$ J/(kg·K), thermal conductivity $\kappa = 16$ W/(m·K). In the 1D equivalent model, we have $\rho_1 c_{p1} = 1.01 \times 10^7$ J/(K·m$^3$), $\rho_2 c_{p2} = 4 \times 10^6$ J/(K·m$^3$), $\kappa_1 = 30$ W/(m·K), $\kappa_2 = 16$ W/(m·K), $b_1 = b_3 = 6.68$ mm, $b_2 = 9.32 - c$, and $b_4 = 9.32 + c$. We take $N + 1 = 20$ units to form a chain and apply constant temperature boundary conditions $T_0 = 293.15$ K to its both ends.

In the experiment, we used stainless steel and made the experimental samples by 3D metal printing. The density, heat capacity, and thermal conductivity of stainless steel are the same as those used in the simulation. To facilitate 3D printing and experimental observation, the chain models were made into rectangles and welded to 16 cm × 8 cm metal plates as bases. The base makes the boundary of the model in better contact with a 16 cm × 8 cm heat sink and ensures stable boundary conditions. The heat sink is made of aluminum and connected with liquid circulation temperature control devices. In the experiment, the temperature of the heat sink were set to room temperature which is about 288 K. The temperature of the hot air gun was over 380 K, which ensures that the sphere can be heated to 40 K above the room temperature. The temperature fields were measured with an infrared thermal camera Fotric 347 with a wide-angle lens and a noise-equivalent temperature difference NETD < 0.04 K.

**References**

1. Hao, M., Li, J., Park, S., Moura, S. & Dames, C. Efficient thermal management of Li-ion batteries with a passive interfacial thermal regulator based on a shape memory alloy. *Nat. Energy* **3**, 899–906 (2018).
2. Nakamura, Y., Sakai, Y., Azuma, M. & Ohkoshi, S. Long-term heat-storage ceramics absorbing thermal energy from hot water. *Sci. Adv.* **6**, eaaz5264 (2020).
3. Kaur, S., Raravikar, N., Helms, B. A., Prasher, R. & Ogletree, D. F. Enhanced thermal transport at covalently functionalized carbon nanotube array interfaces. *Nat. Commun.* **5**, 3082 (2014).
4. Yin, X., Yang, R., Tan, G. & Fan, S. Terrestrial radiative cooling: Using the cold universe as a renewable and sustainable energy source. *Science* **370**, 786–791 (2020).
5. Li, Y. *et al*. Transforming heat transfer with thermal metamaterials and devices. *Nat. Rev. Mater.* **6**, 488–507 (2021).
6. Yang, S., Wang, J., Dai, G., Yang, F. & Huang, J. Controlling macroscopic heat transfer with thermal metamaterials: Theory, experiment and application. *Phys. Rep.* **65** (2021).
7. Narayana, S. & Sato, Y. Heat Flux Manipulation with Engineered Thermal Materials. *Phys. Rev. Lett.* **108**, 214303 (2012).

8. Schittny, R., Kadic, M., Guenneau, S. & Wegener, M. Experiments on transformation thermodynamics: molding the flow of heat. *Phys. Rev. Lett*. **110**, 195901 (2013).

9. Li, Y. *et al*. Temperature-dependent transformation thermotics: from switchable thermal cloaks to macroscopic thermal diodes. *Phys. Rev. Lett.* **115**, 195503 (2015).

10. Li, Y., Li, J., Qi, M., Qiu, C.-W. & Chen, H. Diffusive nonreciprocity and thermal diode. *Phys. Rev. B* **103**, 014307 (2021).

11. Li, Y., Bai, X., Yang, T., Luo, H. & Qiu, C.-W. Structured thermal surface for radiative camouflage. *Nat. Commun*. **9**, 273 (2018).

12. Peng, Y.-G., Li, Y., Cao, P.-C., Zhu, X.-F. & Qiu, C.-W. 3D Printed Meta-Helmet for Wide-Angle Thermal Camouflages. *Adv. Funct. Mater.* **30**, 2002061 (2020).

13. Hu, R. *et al.* Thermal camouflaging metamaterials. *Mater. Today* **45**, 120–141 (2021).

14. Li, J. *et al.* A Continuously Tunable Solid-Like Convective Thermal Metadevice on the Reciprocal Line. *Adv. Mater.* **32**, 2003823 (2020).

15. Xu, G. *et al.* Tunable analog thermal material. *Nat. Commun.* **11**, 6028 (2020).

16. Li, Y. *et al.* Anti–parity-time symmetry in diffusive systems. *Science* **364**, 170–173 (2019).

17. Cao, P, *et al.* High-order exceptional points in diffusive systems: robust APT symmetry against perturbation and phase oscillation at APT symmetry breaking. *ES Energy Environ.* (2019).

18. Xu, L. Geometric phase, effective conductivity enhancement, and invisibility cloak in thermal convection-conduction. *Int. J. Heat Mass Transf.* **11** (2021).

19. Kawabata, K., Shiozaki, K., Ueda, M. & Sato, M. Symmetry and topology in non-Hermitian physics. *Phys. Rev. X* **9**, 041015 (2019).

20. Bender, C. M. Making sense of non-Hermitian Hamiltonians. *Rep. Prog. Phys.* **70**, 947–1018 (2007).

21. Gao, T. *et al.* Observation of non-Hermitian degeneracies in a chaotic exciton-polariton billiard. *Nature* **526**, 554–558 (2015).

22. Leykam, D., Bliokh, K. Y., Huang, C., Chong, Y. D. & Nori, F. Edge modes, degeneracies, and topological numbers in non-Hermitian systems. *Phys. Rev. Lett.* **118**, 040401 (2017).

23. Gong, Z. *et al.* Topological phases of non-Hermitian systems. *Phys. Rev. X* **8**, 031079 (2018).

24. Regensburger, A. et al. Parity–time synthetic photonic lattices. *Nature* **488**, 167–171 (2012).

25. Chong, Y. D., Ge, L., Cao, H. & Stone, A. D. Coherent perfect absorbers: time-reversed lasers. *Phys. Rev. Lett*. **105**, 053901 (2010).

26. Bisson, J.-F. & Nonguierma, Y. C. Single-mode lasers using parity-time-symmetric polarization eigenstates. *Phys. Rev. A* **102**, 043522 (2020).

27. Hasan, M. Z. & Kane, C. L. Colloquium: Topological insulators. *Rev. Mod. Phys.* **82**, 3045–3067 (2010).

28. Heeger, A. J., Kivelson, S., Schrieffer, J. R. & Su, W.-P. Solitons in conducting polymers. *Rev. Mod. Phys.* **60**, 781–850 (1988).

29. Yariv, A. & Yeh, P. *Optical Waves in Crystals: Propagation and Control of Laser Radiation.* (Wiley, 1984).

30. Coulais, C., Fleury, R. & van Wezel, J. Topology and broken Hermiticity. *Nat. Phys.* **17**, 9–13 (2021).

31. Hirsbrunner, M. R., Philip, T. M. & Gilbert, M. J. Topology and observables of the non-Hermitian Chern insulator. *Phys. Rev. B* **100**, 081104 (2019).

32. Kunst, F. K., Edvardsson, E., Budich, J. C. & Bergholtz, E. J. Biorthogonal bulk-boundary correspondence in non-Hermitian systems. *Phys. Rev. Lett.* **121**, 026808 (2018).

33. Yao, S. & Wang, Z. Edge states and topological invariants of non-Hermitian systems. *Phys. Rev. Lett.* **121**, 086803 (2018).

34. Claes, J. & Hughes, T. L. Skin effect and winding number in disordered non-Hermitian systems. *Phys. Rev. B* **103**, L140201 (2021).

35. Lee, T. E. Anomalous edge state in a non-Hermitian lattice. *Phys. Rev. Lett.* **116**, 133903 (2016).

36. Okuma, N., Kawabata, K., Shiozaki, K. & Sato, M. Topological origin of non-Hermitian skin effects. *Phys. Rev. Lett*. **124**, 086801 (2020).

37. Zak, J. Berry's phase for energy bands in solids. *Phys. Rev. Lett.* **62**, 2747–2750 (1989).

38. Delplace, P., Ullmo, D. & Montambaux, G. Zak phase and the existence of edge states in graphene. *Phys. Rev. B* **84**, 195452 (2011).

39. Xiao, M., Zhang, Z. Q. & Chan, C. T. Surface impedance and bulk band geometric phases in one-dimensional systems. *Phys. Rev. X* **4**, 021017 (2014).

40. Kohn, W. Analytic Properties of Bloch Waves and Wannier Functions. *Phys. Rev.* **115**, 809–821 (1959).

41. Martí-Sabaté, M. & Torrent, D. Absence of Quantization of Zak's Phase in One-Dimensional Crystals. arXiv:2107.10144 [cond-mat, physics:physics] (2021).

42. Yoshida, T. & Hatsugai, Y. Bulk-edge correspondence of classical diffusion phenomena. *Sci. Rep.* **11**, 888 (2021).

**Acknowledgements**

C.W.Q acknowledges the support from Ministry of Education, Republic of Singapore, via the grant No.: R-263-000-E19-114. The work at Zhejiang University was sponsored by the National Natural Science Foundation of China (NNSFC) under Grants No. 61625502, No.11961141010, and No. 61975176, the Top-Notch Young Talents Program of China, and the Fundamental Research Funds for the Central Universities.

**Author contributions**

Y.L. conceived the idea and constructed the theory. Y.L., M.Q., and D.W. performed the theoretical analysis and calculation. M.Q. performed the numerical simulations. Y.L., M.Q., and D.W. designed and performed the experiments. Y.L. and M.Q. wrote the manuscript. All the authors contributed to the manuscript editing. Y.L., C.W.Q., and H.C. supervised the work.

**Competing financial interests**

The authors declare no competing financial interests.

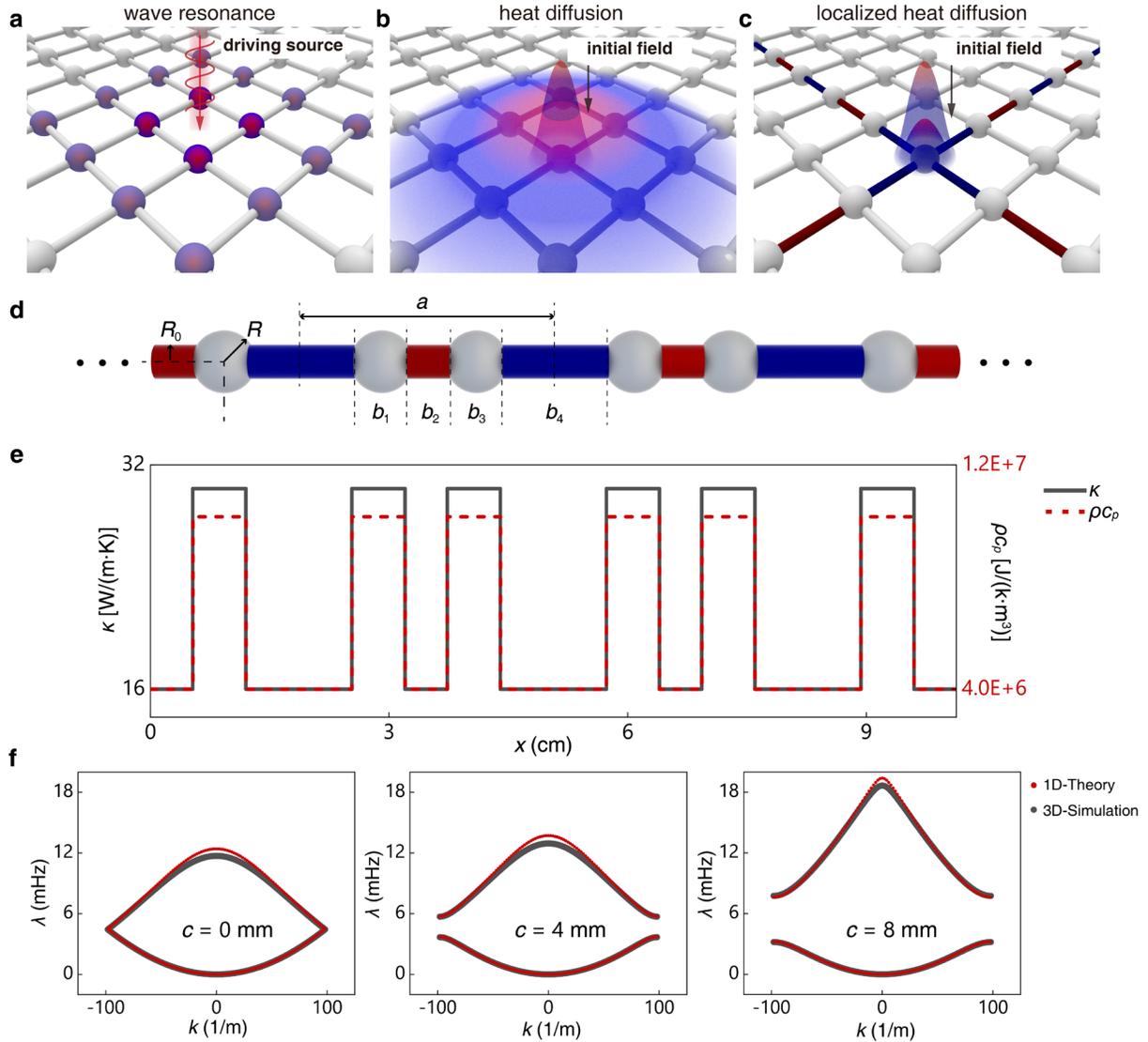

**Figure 1. Heat diffusion in thermal lattices. a**, A wave mode localized on the units can be excited by a driving source at the resonant frequency. **b**, Heat diffusion mode cannot be selectively excited. The temperature field is mainly composed of extended modes and often spreads out. **c**, Conceptual figure of a localized heat diffusion. **d**, A 3D model of spheres and rods. The inter-cell and intra-cell rods are blue and red colored, respectively. **e**, The distributions of thermal conductivity $\kappa$, density $\rho$ and heat capacity $c_p$ in a 1D equivalent model. **f**, The band structures of the system when $c$ = 0 mm, 4 mm, and 8 mm.

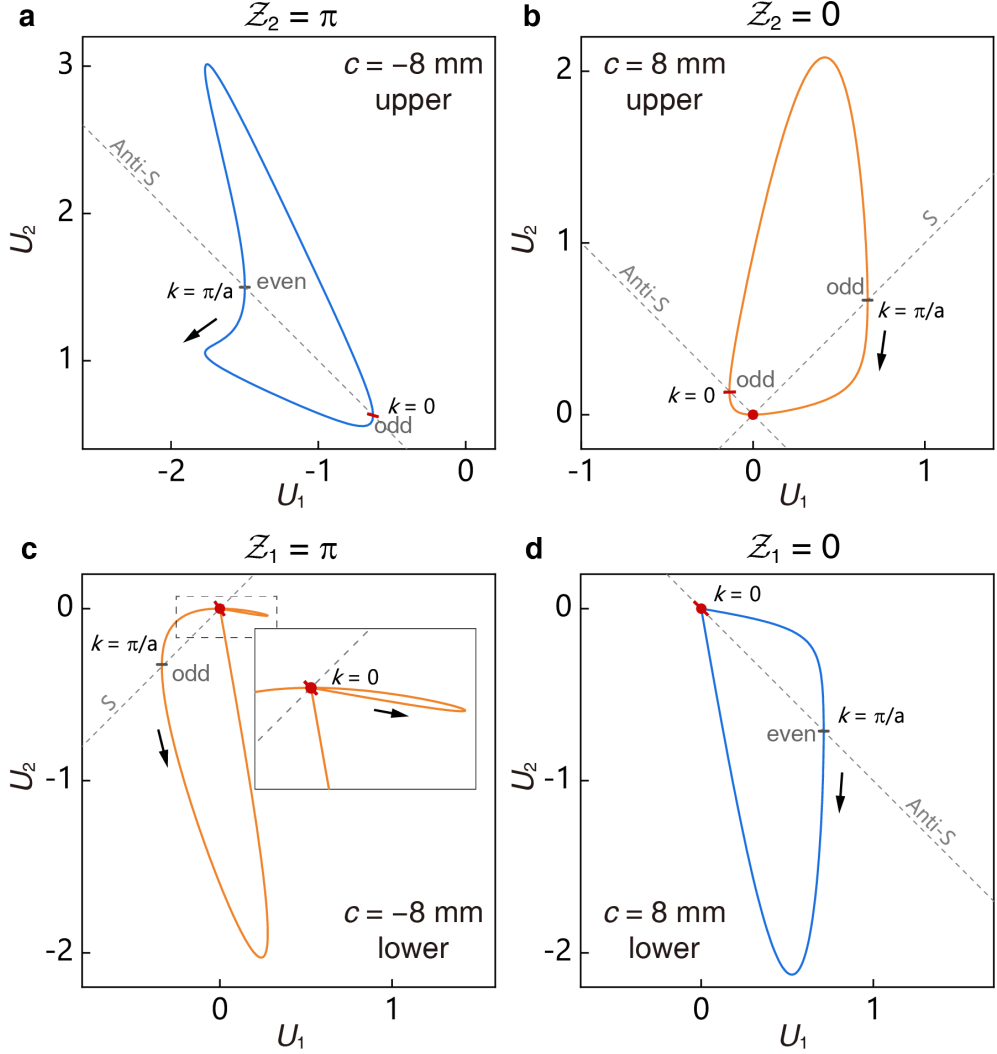

**Figure 2. Trajectories of the state vector.** The trajectories of $U$ as $k$ is increased from $-\pi/a$ to $\pi/a$ in the upper (**a**,**b**) and lower (**c**,**d**) bands for $c = -8$ mm (**a**,**c**) and 8 mm (**b**,**d**). The orange-colored trajectories in **b** and **c** pass the zero point (red dot) with a sign change of $U_1$. The Zak phase of each band is marked on top of the sub-figure. The points where $k = \pm\pi/a$ ($k = 0$) are marked by short grey (red) bars. The grey dashed lines represent states that are symmetric (S) or anti-symmetric (Anti-S) under spatial inversion around the center of the inter-cell rod. The parity (even or odd) of the state is marked at $k = 0$ and $\pm\pi/a$.

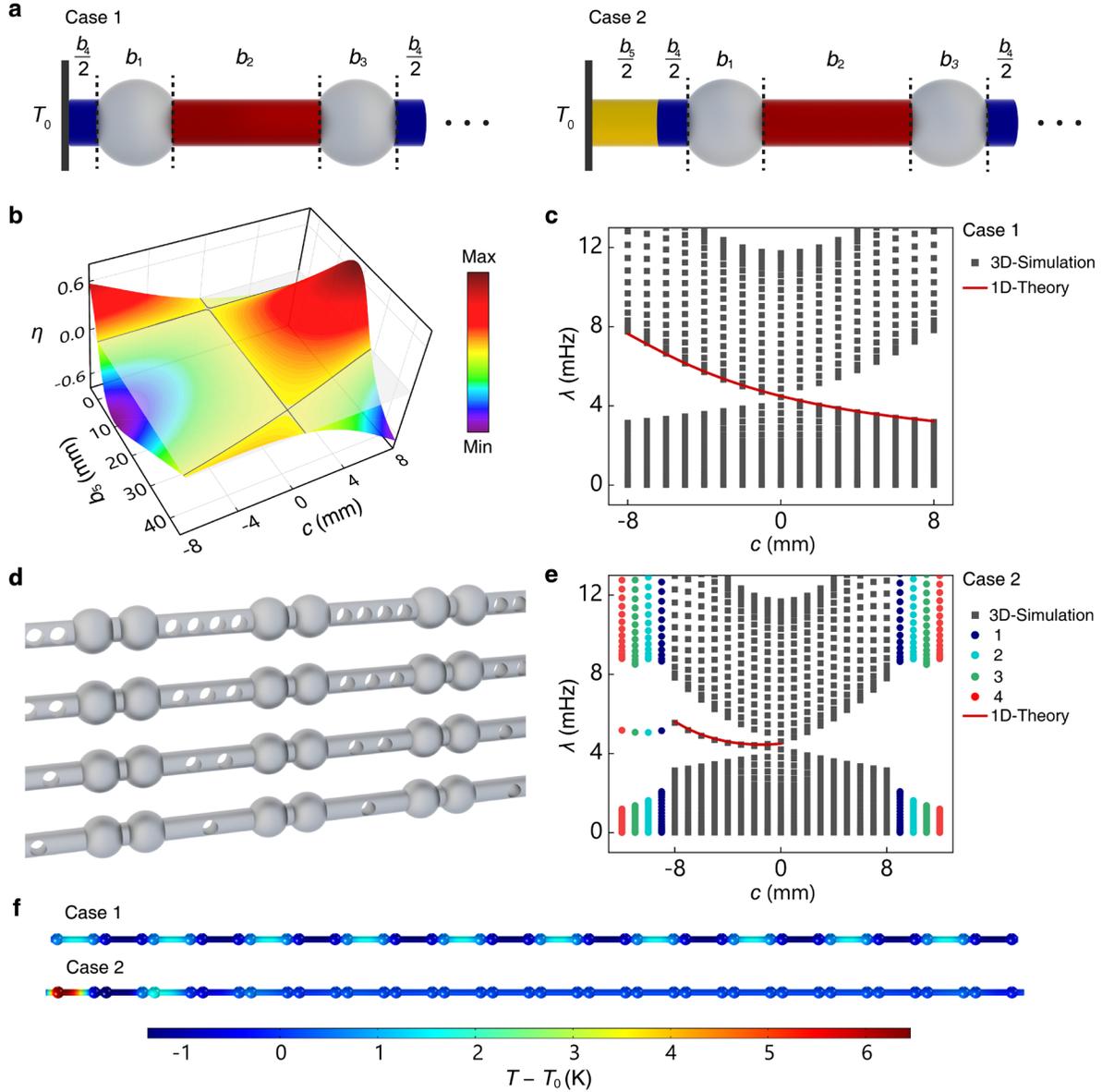

**Figure 3. Influence of boundary condition and optimization of the edge state. a**, Two cases of different boundary selection with constant temperature boundary $T_0$. The yellow part $b_5$ represents the extended part. **b**, The three-dimensional diagram of $\eta$. An edge state exists below the zero plane. **c**, The eigenvalue distribution for Case 1. **d**, The optimized models with 1 to 4 holes on the rods. **e**, The eigenvalue distribution for Case 2. **f**, The temperature distributions of states in or at the edge of the bandgap for Case 1 and Case 2 with $c = -8$ mm.

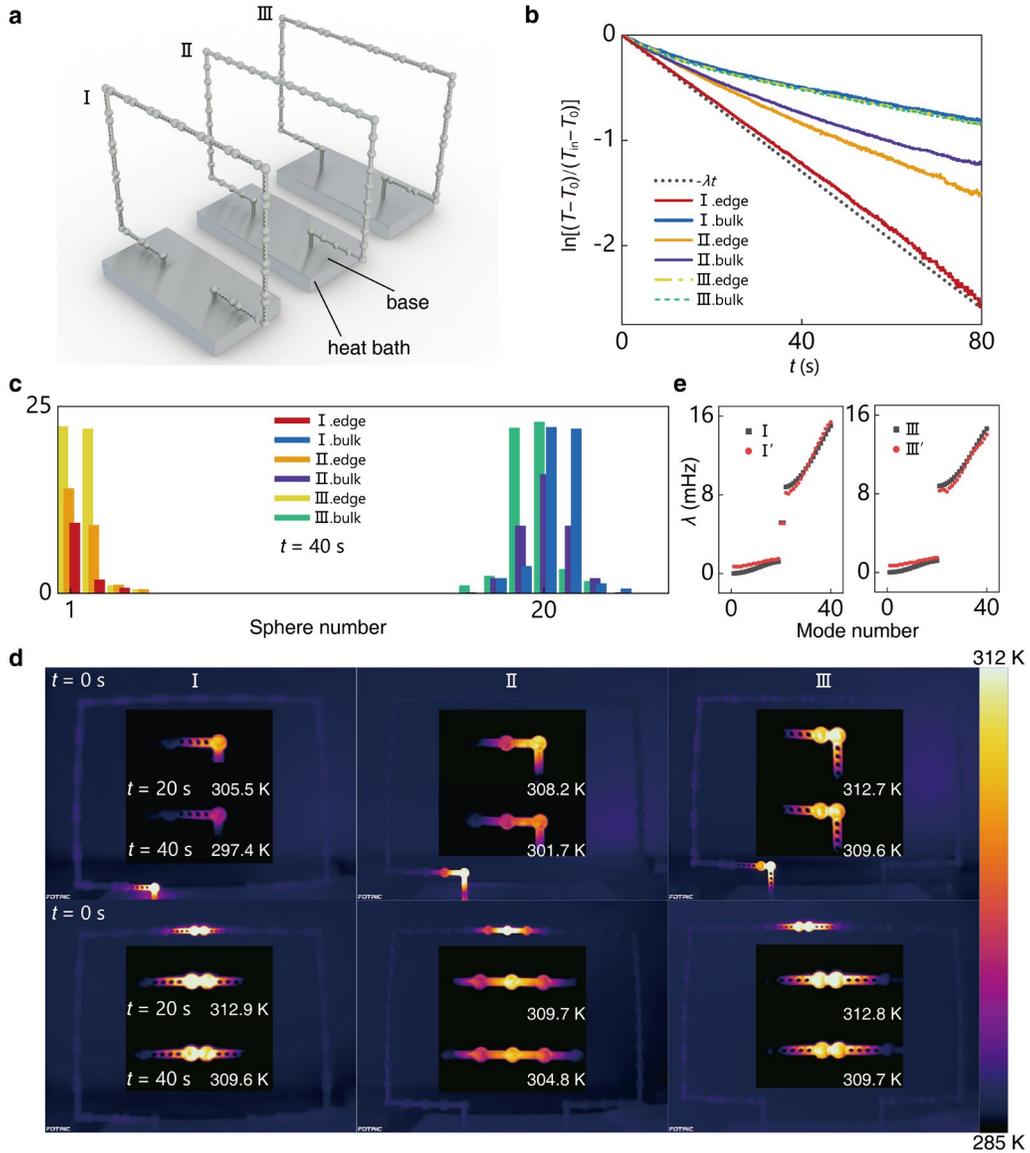

**Figure 4. Experimental setup and results. a**, Physical models of I, II, III made by stainless steel. **b**, The maximum temperature of the models decreasing with time. **c**, The temperature distributions of the spheres $t = 40$ s. **d**, The thermal images of the edge state and the bulk state of the models at $t = 0, 20,$ and $40$ s. The overall pictures represent the temperature distributions at $t = 0$ s, and the insets represent the close-up temperature distributions at $t = 20$ and $40$ s. The temperature values represent the highest temperature of the heated spheres. **e**, Comparison of measured eigenvalues (black) and simulated eigenvalues (red) of model I and III.

**Supplementary Note 1**

**One-dimensional model**

In the 1D model, a unit of the sphere-and-rod model is divided into five regions and the states are described as a piecewise function $u(x)$ in Eq. (4) of the main text.

The matching conditions at the interfaces between the five regions are

$$\begin{cases} u(\zeta_j^-) = u(\zeta_j^+) \\ \kappa(\zeta_j^-) u'(\zeta_j^-) = \kappa(\zeta_j^+) u'(\zeta_j^+) \end{cases} \quad (1)$$

According to the matching conditions, the following relationship can be obtained

$$\begin{pmatrix} e^{ik_1\zeta_{j+1}} & e^{-ik_1\zeta_{j+1}} \\ \kappa_1 k_1 e^{ik_1\zeta_{j+1}} & -\kappa_1 k_1 e^{-ik_1\zeta_{j+1}} \end{pmatrix} \begin{pmatrix} X^{(j+1)}_n \\ Y^{(j+1)}_n \end{pmatrix} = \begin{pmatrix} e^{ik_2\zeta_{j+1}} & e^{-ik_2\zeta_{j+1}} \\ \kappa_2 k_2 e^{ik_2\zeta_{j+1}} & -\kappa_2 k_2 e^{-ik_2\zeta_{j+1}} \end{pmatrix} \begin{pmatrix} X^{(j)}_n \\ Y^{(j)}_n \end{pmatrix} \quad (2)$$

for $j = -2, -1, 0, 1$. And

$$\begin{pmatrix} e^{-ik_2 a} & 0 \\ 0 & e^{ik_2 a} \end{pmatrix} \begin{pmatrix} X^{(-2)}_{n+1} \\ Y^{(-2)}_{n+1} \end{pmatrix} = \begin{pmatrix} X^{(2)}_n \\ Y^{(2)}_n \end{pmatrix} \quad (3)$$

After derivation, the transfer equation $M$ can be obtained

$$\begin{pmatrix} X^{(-2)}_n \\ Y^{(-2)}_n \end{pmatrix} = e^{ika} \begin{pmatrix} X^{(-2)}_{n-1} \\ Y^{(-2)}_{n-1} \end{pmatrix} = \begin{pmatrix} M_{11} & M_{12} \\ M_{21} & M_{22} \end{pmatrix} \begin{pmatrix} X^{(-2)}_{n-1} \\ Y^{(-2)}_{n-1} \end{pmatrix} \quad (4)$$

where $M_{11}$ and $M_{12}$ are given by

$$\begin{cases} M_{11} = \cos(2\alpha)\cos(2\beta) - \cosh\gamma \sin(2\alpha)\sin(2\beta) + 2\sinh^2\gamma \sin^2\alpha \sin(\beta+\delta)\sin(\beta-\delta) \\ \quad + i[\cos(2\alpha)\sin(2\beta) + \cosh\gamma \sin(2\alpha)\cos(2\beta) - 2\sinh^2\gamma \sin^2\alpha \sin(\beta-\delta)\cos(\beta+\delta)] \\ M_{12} = 2ie^{ik_2 a} \sinh\gamma \sin\alpha [\cos\alpha \cos(\beta-\delta) - \cosh\gamma \sin\alpha \sin(\beta-\delta)] \end{cases} \quad (5)$$

Also, $M_{22} = M_{11}^*$ and $M_{21} = M_{12}^*$. For $c = 0$, the two rods have the same lengths, $b_2 = b_4$.

$$\text{Re}(M_{11}) = 2(\cosh\gamma \sin\alpha \sin\beta - \cos\alpha \cos\beta)^2 - 1 \quad (6)$$

According to Eq. S(4), the eigenvalue of $M$ should be

$$e^{ika} = \frac{1}{2}(M_{11} + M_{22}) \pm i\sqrt{(M_{11}M_{22} - M_{12}M_{21}) - \left(\frac{M_{11} + M_{22}}{2}\right)^2} \quad (7)$$

We thus have

$$\cos ka = \frac{M_{11} + M_{22}}{2} = \text{Re}(M_{11}) = 2\left(\cosh\gamma \sin\alpha \sin\beta - \cos\alpha \cos\beta\right)^2 - 1 \quad (8)$$

The gap is closed at $k = \pm\pi/a$, where

$$\cot\alpha \cot\beta = \cosh\gamma \quad (9)$$

For $c \neq 0$, the two rods have different lengths, $b_2 \neq b_4$. A bandgap exists in this case. Its upper and lower edges are reached at $k = \pm\pi/a$, where

$$\sin\beta\left(\cosh\gamma - \cot\alpha \cot\beta\right) = \pm\sinh\gamma \sin\delta \quad (10)$$

**Supplementary Note 2**

**SSH model results**

We also consider using the SSH model to discretized the heat transfer lattice system. Mark the left and right spheres in a unit as $A$ and $B$. $T_{An}$ and $T_{Bn}$ are the average temperatures of spheres $A$ and $B$, respectively. According to Fourier's law in heat conduction, the rate of internal energy change of each sphere is equal to the heat flow in and out as

$$\rho c_p V_{\text{sphere}} \frac{\partial T_{A,Bn}}{\partial t} = \Phi_{\text{left}} - \Phi_{\text{right}} \quad (11)$$

where $\Phi_{\text{left}}$ is the heat flow into the sphere from its left boundary, $\Phi_{\text{right}}$ is the heat flow out of the sphere from its right boundary.

Assuming that the temperature distribution on the rod is linear, the heat flow can be approximated as proportional to the temperature difference

$$\Phi = -\pi R_0^2 \kappa \frac{dT}{dx} \approx -\pi R_0^2 \kappa \frac{T_{\text{right}} - T_{\text{left}}}{b_{2,4}} \quad (12)$$

where $T_{\text{right}}$ and $T_{\text{left}}$ are the temperatures at the right and left ends of the rod. We approximate $T_{\text{right}} \approx T_{Bn}$ and $T_{\text{left}} \approx T_{An}$ for intra-cell, and $T_{\text{right}} \approx T_{An}$ and $T_{\text{left}} \approx T_{Bn}$ for inter-cell. A discretized

approximation version of Eq. S(11) is thus

$$\begin{cases} \rho c V_{sphere} \dfrac{\partial T_{An}}{\partial t} = \pi R_0^2 \kappa \left( \dfrac{T_{Bn-1} - T_{An}}{b_4} + \dfrac{T_{Bn} - T_{An}}{b_2} \right) \\ \rho c V_{sphere} \dfrac{\partial T_{Bn}}{\partial t} = \pi R_0^2 \kappa \left( \dfrac{T_{An} - T_{Bn}}{b_2} + \dfrac{T_{An+1} - T_{Bn}}{b_4} \right) \end{cases} \quad (13)$$

where $V_{sphere}$ is approximated by $4\pi R^3/3$. The inter-cell coupling coefficient $\tau_1$ and the intra-cell coupling coefficient $\tau_2$ can be obtained

$$\begin{cases} \tau_1 = \dfrac{3R_0^2}{4R^3 b_2} D \\ \tau_2 = \dfrac{3R_0^2}{4R^3 b_4} D \end{cases} \quad (14)$$

where $D = \kappa/\rho c_p$ is the diffusivity.

The Hamiltonian for this diffusive SSH model is

$$H_{SSH}(k) = -i \begin{pmatrix} \tau_1 + \tau_2 & -\tau_1 - \tau_2 e^{-ika} \\ -\tau_1 - \tau_2 e^{ika} & \tau_1 + \tau_2 \end{pmatrix} \quad (15)$$

where $k$ is the wavenumber. Setting $\Delta = \tau_1/\tau_2$, the spectrum of the system can be obtained as

$$\omega(k) = -i\tau_2 \left[ (1+\Delta) \pm \sqrt{(\Delta + \cos k)^2 + \sin^2 k} \right] \quad (16)$$

The eigenvalues of the system are predicted by the discrete Hamiltonian matrix of the SSH model. However, the SSH model is only approximately valid under certain conditions because the actual temperature field is nonlinear in the sphere-and-rod model. It can be found that the SSH model's prediction is quite different from the simulation results as shown in Fig. S1.

**Supplementary Note 3**

**Influence of the boundary conditions**

The appearance of the boundary state is affected by the imposed boundary conditions. For the 3D model Case 2 in Fig. 3a of the main text, gradually increase the extended boundary part

$b_5$. Then the eigenvalues will gradually move from the upper edge of the upper band to the lower band until it falls into the edge of the lower band when $c = -8$ mm. It can be seen that when the extended part $b_5 = (b_1 + b_4) = 8$ mm, the eigenvalue moves to the middle of the gap (Fig. S3).

At the same time, under this condition, if the constant temperature condition at the boundary is replaced by thermally insulated condition, we have

$$\begin{cases} X^{(-2)}_0 e^{-ik_2(a+b_5)/2} - Y^{(-2)}_0 e^{ik_2(a+b_5)/2} = 0 \\ X^{(2)}_N e^{ik_2(a+b_5)/2} - Y^{(2)}_N e^{-ik_2(a+b_5)/2} = 0 \end{cases} \quad (17)$$

As $\Lambda_+$ approaches zero for large $N$, we have

$$X^R_- X^L_- e^{ik_2 b_5} - Y^R_- Y^L_- e^{-ik_2 b_5} + X^R_- Y^L_- e^{-ik_2 a} - X^L_- Y^R_- e^{ik_2 a} = 0 \quad (18)$$

It follows that

$$\begin{cases} \sqrt{\text{Re}(M_{11})^2 - 1} = U_1 \sin k_2 b_5 \\ \text{Im}(M_{11}) = -U_1 \cos k_2 b_5 \end{cases} \quad (19)$$

where the requirement $\sqrt{\text{Re}(M_{11})^2 - 1} > 0$ is satisfied when $c > 0$, and the boundary states of the system will appear on the side of $c > 0$ as shown in Fig. S4.

We can further generalize the discussion to hybrid boundary conditions. For example, it is common that the system is coupled with the environment through the boundaries with a heat-exchange rate $h$. In that case,

$$\begin{cases} EX^{(-2)}_0 e^{-ik_2(a+b_5)/2} + E^* Y^{(-2)}_0 e^{ik_2(a+b_5)/2} = 0 \\ EX^{(2)}_N e^{ik_2(a+b_5)/2} + E^* Y^{(-2)}_N e^{-ik_2(a+b_5)/2} = 0 \end{cases} \quad (20)$$

where $E = h + i\kappa_2 k_2$. It follows that

$$|E|^2 (X^R_- X^L_- e^{ik_2 b_5} - Y^R_- Y^L_- e^{-ik_2 b_5}) - E^2 e^{-ik_2 a} X^R_- Y^L_- + E^{*2} e^{ik_2 a} Y^R_- X^L_- = 0 \quad (21)$$

or

$$U_1 \begin{pmatrix} \sin k_2 b_5 \\ \cos k_2 b_5 \end{pmatrix} = \mathbf{R}(2\theta) \begin{pmatrix} \sqrt{\operatorname{Re}(M_{11})^2 - 1} \\ \operatorname{Im}(M_{11}) \end{pmatrix} \quad (22)$$

where $\mathbf{R}(2\theta)$ is the rotation matrix of $2\theta$, and $\theta = \operatorname{Arg}(E)$. One may check the equation for constant temperature boundary condition in the main text, and Eq. S(19) for thermally insulated boundary condition are special cases of Eq. S(22) at $\theta = 0$ and $\pi/2$, respectively. The existence of an edge state is thus determined by

$$\eta = U_1 \sin(k_2 b_5 + 2\theta) < 0 \quad (23)$$

Therefore, the bulk-boundary correspondence of the system is still established through the sign of $U_1$ and a phase factor $k_2 b_5 + 2\theta$ characterizing the boundary conditions. As an example, we estimate that $\theta$ is around $\pi/4$ for a heat exchange rate $h = 1600$ W/m². The results are plotted in Fig. S5, showing a good accuracy of the theoretical prediction.

We may also make an interface between the present model with an inverted one as shown in Fig. S5a. In this case, we can observe the interface state of the system under thermally insulated boundary conditions. We set the rod length between the interface as $b_6$. It can be found that the eigenvalue of the interface state appears at the middle of the bandgap when $c = -8$ mm and $b_6 = 9.32$ mm (Fig. S6b).

To further prove our theory, we conducted a time-domain simulation. In simulation, we take temperature boundary conditions $T_0 = 293.15$ K and add the initial temperature $T_{in} = 333.15$ K to the 1st (edge) and 20th (bulk) spheres respectively. The simulation results for the three models in Fig. 4a of the main text could be found in Fig. S7.

We also consider the case when all other spheres were removed except the one at the boundary. After comparison, it can be found that the single-sphere model has a faster temperature dissipation rate than the sphere-and-rod model at the beginning, but after a short time, its temperature dissipation rate is gradually surpassed (Fig. S8).

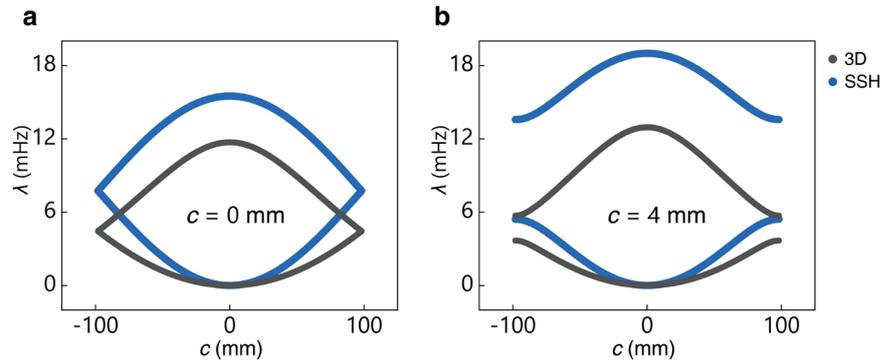

**Supplementary Figure 1. SSH model.** The band distributions of the system when $c = 0$ mm (**a**) and 4 mm (**b**). The SSH prediction and 3D simulation solutions are represented by blue and black lines.

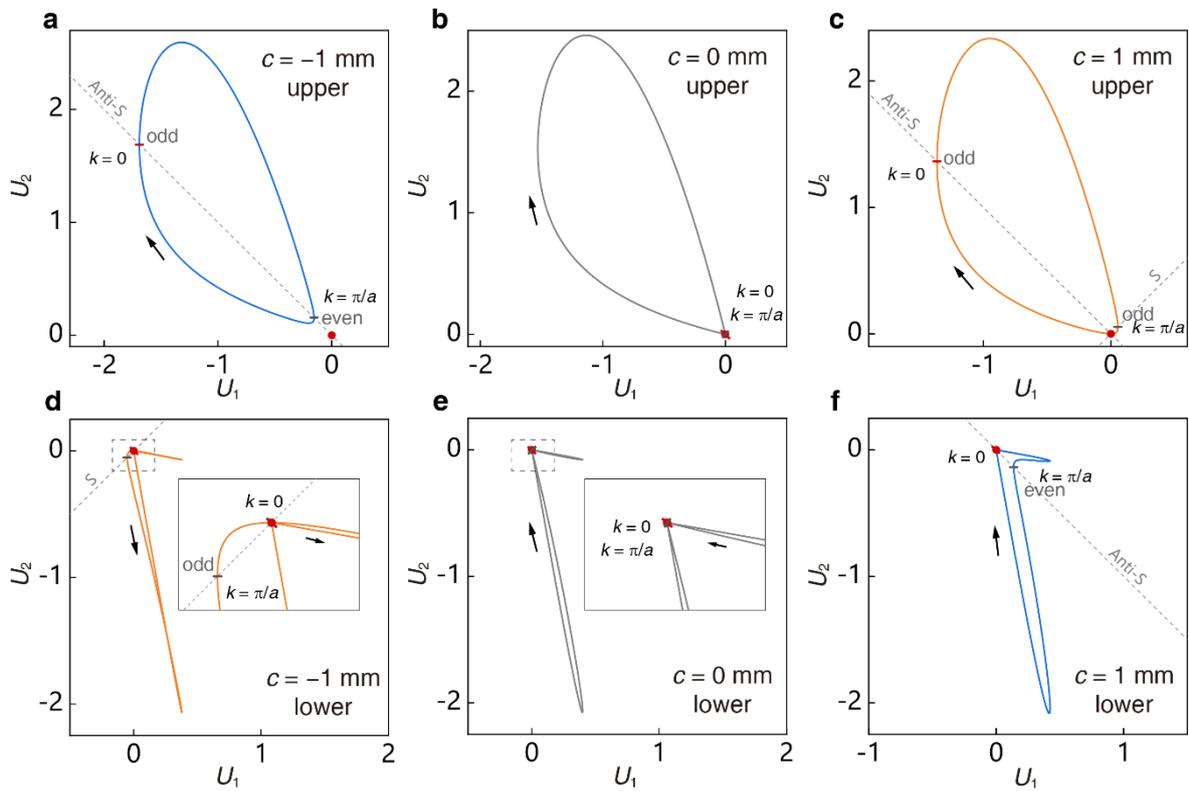

**Supplementary Figure 2. Trajectories of the state vector. a**, Trajectory of the upper band when $c = -0.1$ mm. **b**, Trajectory of the upper band when $c = 0$ mm. **c**, Trajectory of the upper band when $c = 1$ mm. **d**, Trajectory of the lower band when $c = -0.1$ mm. **e**, Trajectory of the lower band when $c = 0$ mm. **f**, Trajectory of the lower band when $c = 1$ mm.

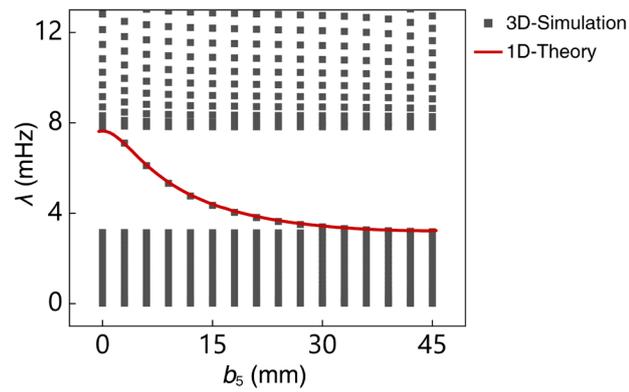

**Supplementary Figure 3. Influence of the extension length.** Eigenvalue distributions with the increase of the length of each extended part $b_5$ when $c = -8$ mm.

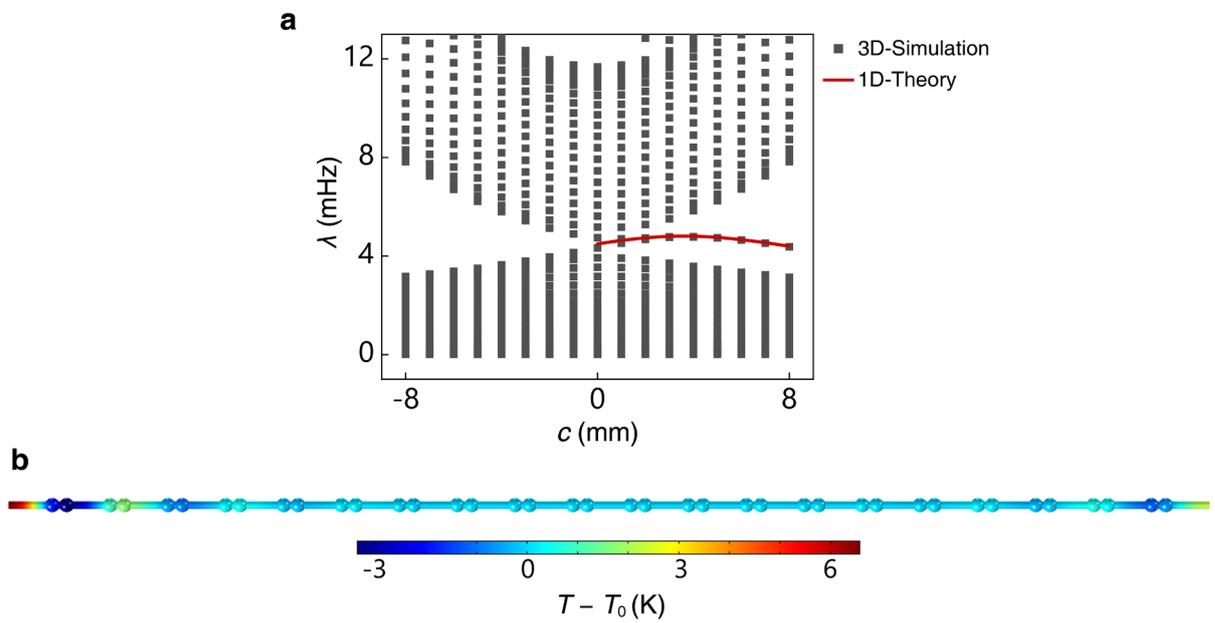

**Supplementary Figure 4. Thermally insulated boundary conditions. a**, Eigenvalue distributions under adiabatic boundary conditions. **b**, Field distribution of the edge state for $c = 8$ mm.

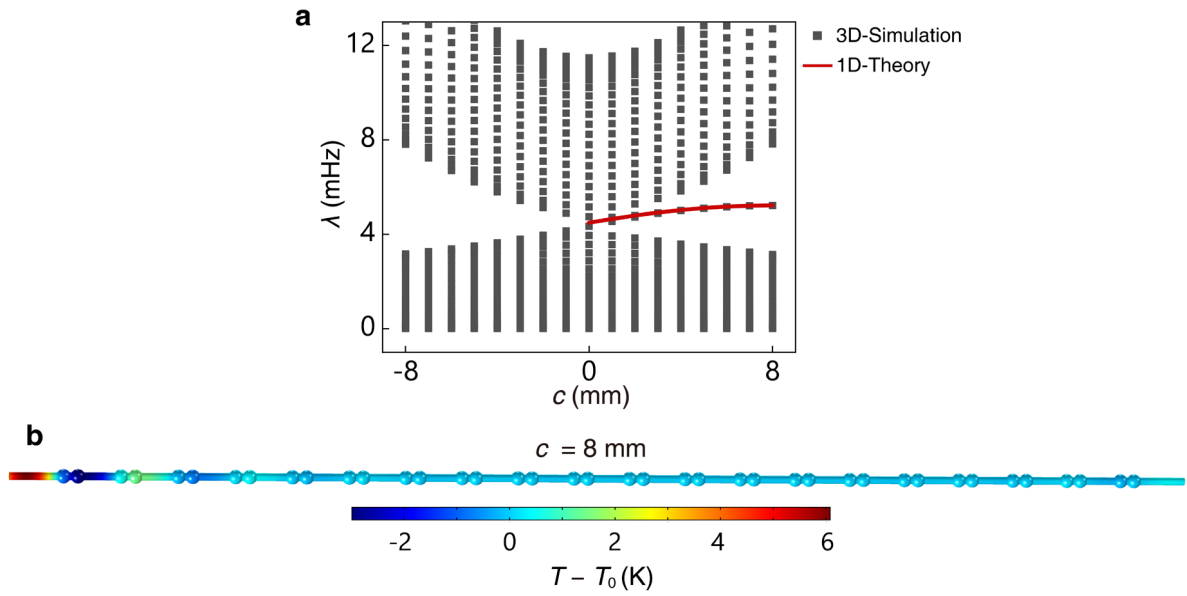

**Supplementary Figure 5. Hybrid boundary conditions. a**, Eigenvalue distributions with extension length $b_5 = 36$ mm and heat exchange rate $h = 1600$ W/m$^2$ at the left and right ends. **b**, Field distribution of the edge state for $c = 8$ mm.

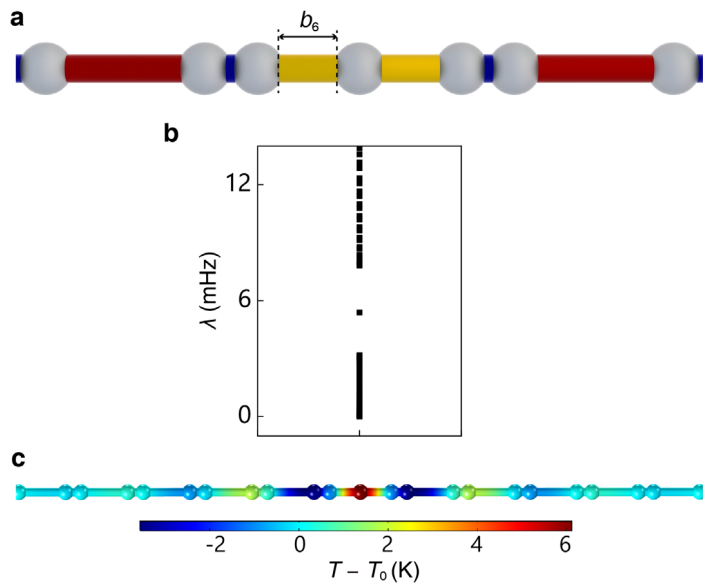

**Supplementary Figure 6. Interface state. a**, Schematic of the model. **b**, Eigenvalue distribution. **c**, The temperature distribution of the interface state.

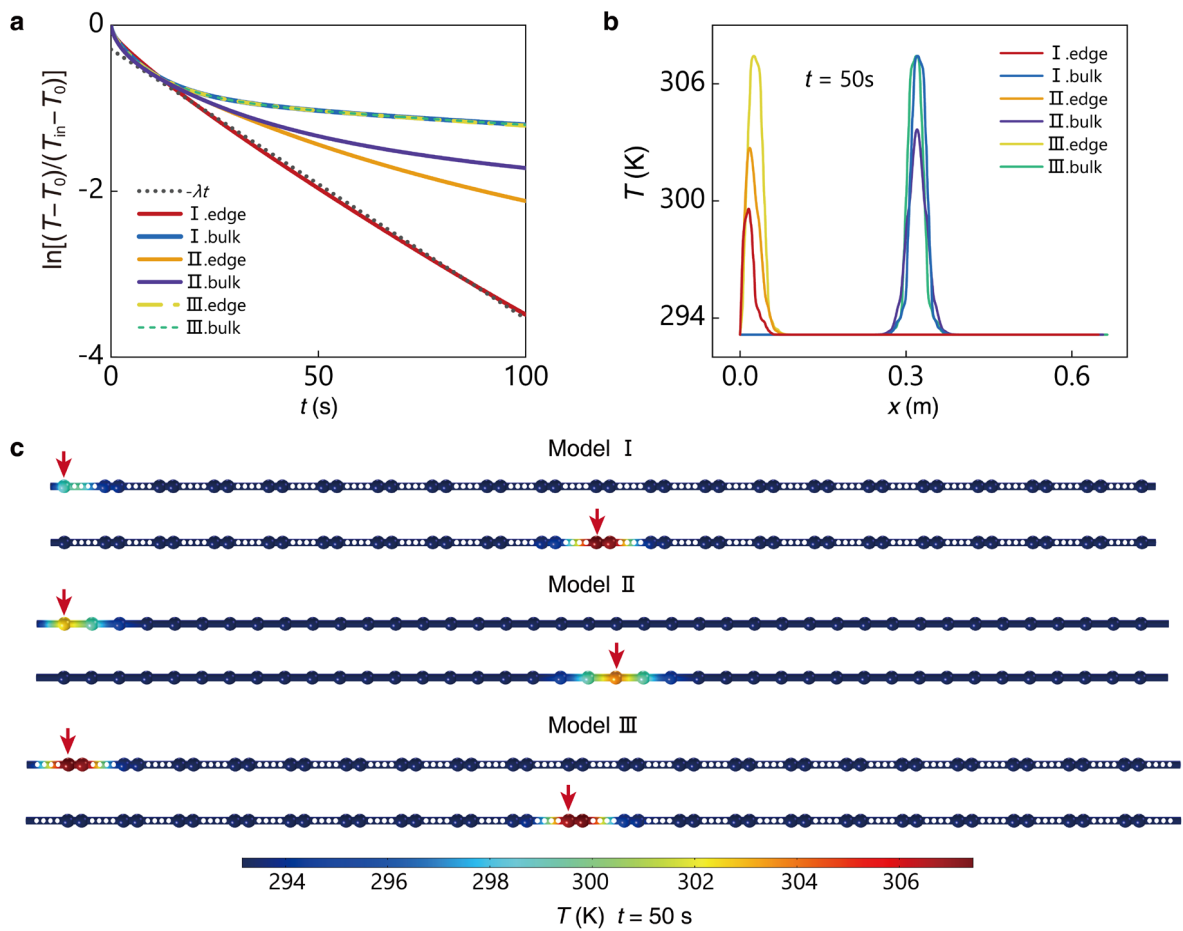

**Supplementary Figure 7. Simulation results of model I, II, and III. a**, The decay rate of maximum temperatures of the model I, II, and III. **b**, The temperature distribution curves of model I, II, and III at $t = 50$ s. **c**, The temperature distribution fileds of model I, II, and III at $t = 50$ s. The positions where the initial temperature is applied are marked by red arrows.

**Supplementary Figure 8. Single-sphere model. a**, Comparison of temperature dissipation rates of the boundary state in model I and the single-sphere model. **b**, Comparison of temperature distributions at $t = 100$ s. **c**, The temperature distribution of the single rod model at $t = 50$ s and 100 s.